\documentclass[reprint,superscriptaddress,
 amsmath,amssymb,
 aps,
 prl,
prl]{revtex4-2}

\usepackage{graphicx}
\usepackage{dcolumn}
\usepackage{bm}
\usepackage{xcolor}

\makeatletter
\newcommand*{\rom}[1]{\expandafter\@slowromancap\romannumeral #1@}

\makeatother

\begin{document}

\title{Acousto-optic modulated vortex beam with tunable topological charge}

\author{A. Pitanti}
\affiliation{Paul-Drude-Institut f{\"u}r Festk{\"o}rperelektronik, Leibniz-Institut im Forschungsverbund Berlin e. V., Hausvogteiplatz 5-7, 10117 Berlin, Germany}
\affiliation{University of Pisa, Dipartimento di Fisica E. Fermi,
largo Bruno Pontecorvo 3, Pisa 56127, Italy}
\altaffiliation[Also at ]{CNR-Istituto Nanoscienze, piazza San Silvestro 12, 56127 Pisa - Italy}
\email[corresponding author: ]{alessandro.pitanti@unipi.it}
\author{N. Ashurbekov}
\affiliation{Paul-Drude-Institut f{\"u}r Festk{\"o}rperelektronik, Leibniz-Institut im Forschungsverbund Berlin e. V., Hausvogteiplatz 5-7, 10117 Berlin, Germany}
\author{I. dePedro-Embid}
\affiliation{Paul-Drude-Institut f{\"u}r Festk{\"o}rperelektronik, Leibniz-Institut im Forschungsverbund Berlin e. V., Hausvogteiplatz 5-7, 10117 Berlin, Germany}
\author{M. Msall}
\affiliation{Paul-Drude-Institut f{\"u}r Festk{\"o}rperelektronik, Leibniz-Institut im Forschungsverbund Berlin e. V., Hausvogteiplatz 5-7, 10117 Berlin, Germany}
\affiliation{Department of Physics and Astronomy, Bowdoin College, Brunswick, Maine 04011, USA}
\author{P. V. Santos}
\email[corresponding author: ]{santos@pdi-berlin.de}
\affiliation{Paul-Drude-Institut f{\"u}r Festk{\"o}rperelektronik, Leibniz-Institut im Forschungsverbund Berlin e. V., Hausvogteiplatz 5-7, 10117 Berlin, Germany}

\date{\today}

\begin{abstract}

Controlling the symmetry of optical and mechanical waves is pivotal to their full exploitation in technological applications and topology-linked fundamental physics experiments. Leveraging on the control of orbital angular momentum, we introduce here a device forming super-high-frequency-modulated optical vortex beams via acousto-optic interaction of light with acoustic vortices. Originated by shape-engineering of a single-contact bulk acoustic wave resonator, we generate vortices in a wide band of frequencies around 4 GHz and with frequency/geometrically tunable topological charge from 1 to 13 and beyond. With all electrical control and on-chip integration our device offers new solutions for angular-momentum-based light communication, three-dimensional particle manipulation, as well as novel interaction schemes for optomechanical devices. 

\end{abstract}

\maketitle

\section{\label{sec:intro}Introduction}

The observation that helical phase fronts of light carry an orbital angular momentum~\cite{Allen1992} ignited a wide research activity focused on harnessing the full vectorial properties of photons, adding spin and orbital angular momentum (SAM/OAM) to the scalar amplitude and phase \cite{Padgett2017,shen2019}. The benefits of angular momentum control include new, high-bandwidth schemes that exploit the OAM basis in optical communications  \cite{Wang2022}, the possibility of trapping and rotational manipulation of particles via optical tweezers \cite{ONeil2002}, higher dimensionality interferometric sensing \cite{Maurer2011}, all of these on top of several fundamental physical effects linked to topology \cite{Parappurath2020,Barik2020} and other applications \cite{shen2019}.

The impressive advancements of vectorial and chiral photonics have been recently mirrored by fascinating results in the investigation of acoustic and elastic waves, which have evolved from the main component of radio-frequency filters \cite{Morgan2010} and oscillators \cite{Bernardo2002} to tools for the manipulation of single excitation in quantum information applications \cite{Delsing2019}. 
The route for achieving a full vectorial control of elastic waves has been mostly focused on a frequency range between tens of Hz and a few MHz \cite{Guo2022}: here the angular momentum-carrying waves can propagate in fluids and have been predominantly used to exert forces on small particles, which can be trapped in three-dimensional potentials and/or subjected to a torque for rotational actuation \cite{Volke2008,Gong2020,Li2021}.

Coupling together vectorial light and vibrations at ultra-high (0.3 - 3 GHz - UHF) and super-high (3 - 30 GHz - SHF) frequency reveals a new class of on-chip, chiral acousto-optic devices, which can unveil several intriguing effects, lead by angular momentum-based modulation of light. These add to the already demonstrated surface acoustic wave amplitude/phase modulators \cite{DeLima2005,pitanti2023gigahertz}, eventually enriching the schemes for OAM multiplexing, which has shown promising results for increased channel transfer capacity with static all-optical vortex generation \cite{wang2012,ren2016,xie2018}.
Time-varying optical vortex beams have recently emerged as higher-dimensional structured light channels \cite{he2022}, spanning from spatio-temporal optical beams with longitudinal \cite{rego2019,jiang2024} or transversal \cite{chong2020,wan2023} time-dependent OAM. The added dimensionality unveils a new potential for space-time differentiators \cite{liu2022,huang2022} and subvelocity of light and superluminal pulse propagation \cite{petrov2019,mazanov2022}.
Conversely, UHF and SHF chiral acoustics provides innovative methods for the topological control of mechanical waves, expanding on concepts of wave manipulation through  acoustic metasurfaces \cite{Zanotto2022, Pitanti2023} and offering new tools for control of hybrid quantum systems as, for example, acoustic exciton polaritons \cite{Santos2023}.

In this manuscript we introduce a novel approach for chiral acousto-optics by manipulating the shape of a piezoelectric bulk acoustic wave resonator (BAWR) to induce the generation of elastic vortex beams within a solid-state substrate. The vortices are generated in a wide frequency band in the GHz range and used to modulate the angular momentum of reflected light, introducing a new approach for the generation of time-varying OAM optical beams.
The vortex topological invariant, also called topological charge ($\ell$), of the acousto-optical vortex beams can be externally controlled in a very wide range in a single device (from $\ell=1$ to 13 and beyond) by simply changing the driving frequency, with the additional option of operating at negative or within an increased $\ell$ range by introducing simple geometrical modifications to the basic device structure.
The $\ell$ tunability, broadband light interaction and high operating frequency (also resulting in a reduced footprint) establish our platform as a new class of versatile acousto-optic devices for hybrid systems, advanced acoustics and light-based telecommunications.
\begin{figure*}[t!]
\includegraphics[width=0.65 \textwidth]{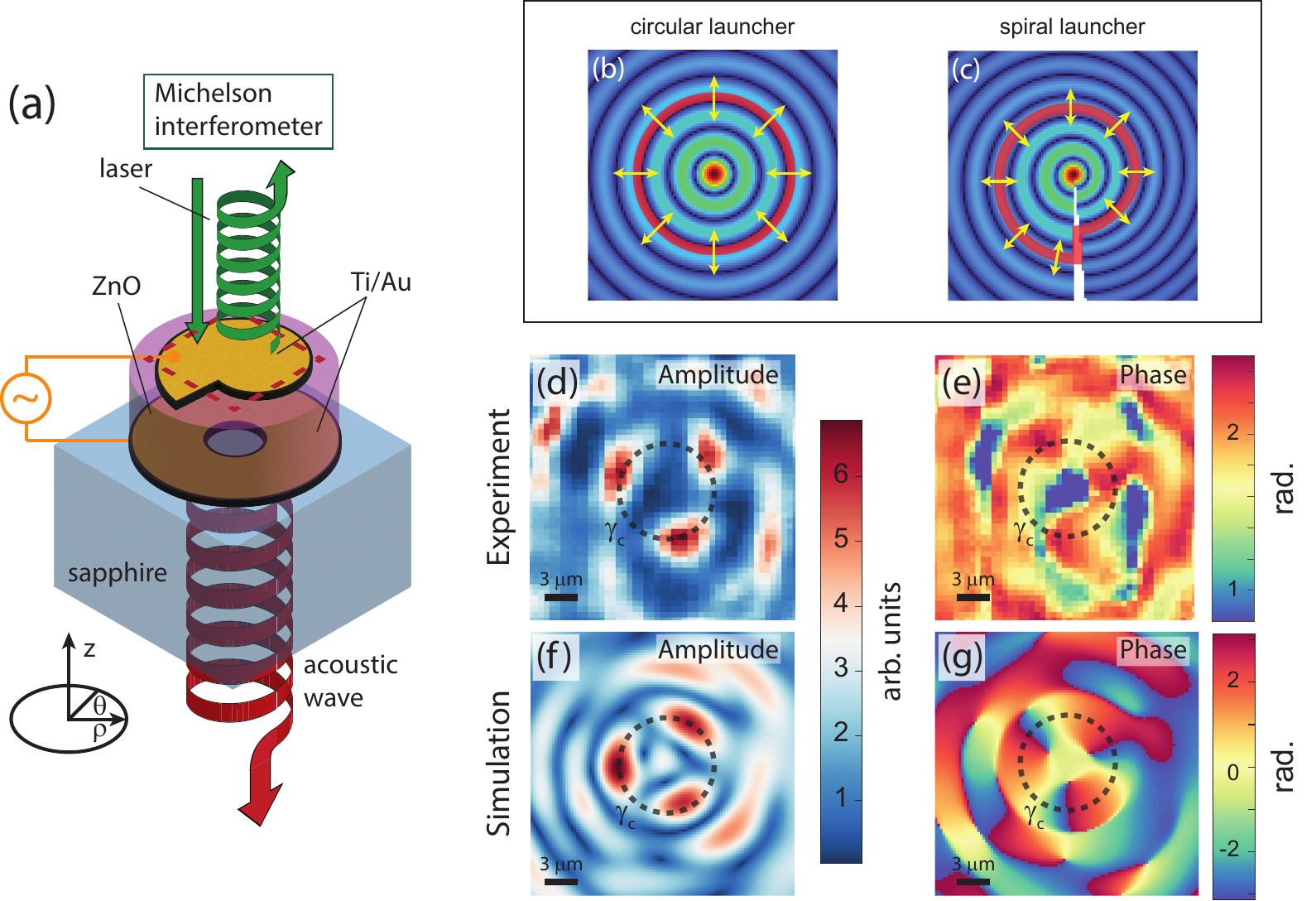}
\caption{\label{fig:1} {\bf Acousto-optical generation of chiral optical beams.} (a): Device concept. A properly engineered bulk acoustic wave resonator (BAWR)  launches an acoustic vortex into the substrate and to modulate in time the OAM of a reflected optical beam. The BAWR is engineered starting from a circular contact (with the expected acoustic field sketched in b) that is perturbed into a spiral (expected field sketched in c). (d-e): Experimental amplitude and phase map of a vortex with $\ell=3$ at 1.139 GHz. (f-g): Simulated maps for the same drive and device geometry extracted from full, 3D simulations.}
\end{figure*}

\section{\label{sec:conc}Device concept}

The basic device concept is displayed in Fig. \ref{fig:1}(a). A bulk acoustic wave resonator (BAWR) composed by a ZnO piezoelectric layer sandwiched between metallic contacts (see Methods for more details) is placed upon a double-polished sapphire substrate. A proper shaping of the BAWR contact is used to generate an acoustic vortex propagating along the substrate; shining light at the device top surface results in the creation of a time-varying OAM reflected optical beam. In our experiment the focused laser light is analyzed by a Michelson interferometer, which allows extraction of coherent field maps in the acousto-optical interaction plane (sketched as a dark red dashed square in figure).

In order to generate vortices, the BAWR contact shape must be modified to produce opportune phase profiles in the $\theta-\rho$ plane; a standard circular launcher, in fact, creates a radial wavefront which can be described, in first approximation, by Bessel's functions of the first kind, $J_l(2\pi\rho/\lambda)$, where $\lambda$ is the radial acoustic wavelength, (see the top-view sketch of Fig. \ref{fig:1}(b)).
Considering the rigorous definition of $\ell$ as \cite{shen2019}:
\begin{equation}
    \ell=\frac{1}{2\pi}\oint_{\gamma_c}\nabla\phi(\vec{r})dr,
\end{equation}
where $\gamma_c$ is a tiny loop surrounding the singularity, it follows that the phase $\phi$ of a standard vortex field with topological charge $\ell$ linearly increases from 0 to $2\pi \ell$ along the path $\gamma_c$; considering an isotropic sound velocity distribution in the $\theta-\rho$ plane, the required phase profile can be obtained by modifying the circular launcher into the shape of an Archimedean Spiral (AS), see the sketch of traveling wavefronts of Fig. \ref{fig:1}(c). The AS is conveniently defined considering the trajectory of a point whose distance from the center linearly increases with the azimuthal angle $\hat{\theta}$ and it has been previously employed to generate vortices in different systems including optical beams~\cite{Kim2010}, hyperbolic polaritons~\cite{Wang2022b}, plasmons~\cite{hachtel2019} and low-frequency sound waves~\cite{Wang2022c}. For practical reasons, including broadband operation, we introduce the AS in the BAWR contact by defining its outer rim as a single spiral loop: 
\begin{equation}\label{eq:spiral}
AS(\theta) = R_0 + g \frac{\theta}{2\pi},\;\text{with}\;\theta\in[0,2\pi]
\end{equation}
where $R_0$ is the spiral starting radius and $R_0 + g$ is its maximum radial extension.

The amplitude and phase maps of a typical vortex generated for an applied driving frequency of 1.14~GHz are shown in Fig.~\ref{fig:1}(d-e). The maps have been obtained using the 532~nm light from a solid state laser; the beam was focused on the device top surface and the amplitude and phase profiles of optics and acoustics analyzed with a Michelson interferometer coupled with a vector-network analyzer for coherent detection of the complex amplitude of the beam (more details can be found in the Methods section).
Finite-element simulations of the same structure within a simplified, albeit 3D toy-model (see Methods) well reproduce the observed feature, both for the amplitude [Fig.~\ref{fig:1}(f)] and phase [Fig.~\ref{fig:1}(g)]. The latter map reproduces qualitatively the same features observed in the experiment although with a scaled range whose origin will become clear in the following discussion. A comparison of experimental and simulated vortex animation is shown as a Supplementary Video.\\

\begin{figure}[t!]
\includegraphics[width=0.45 \textwidth]{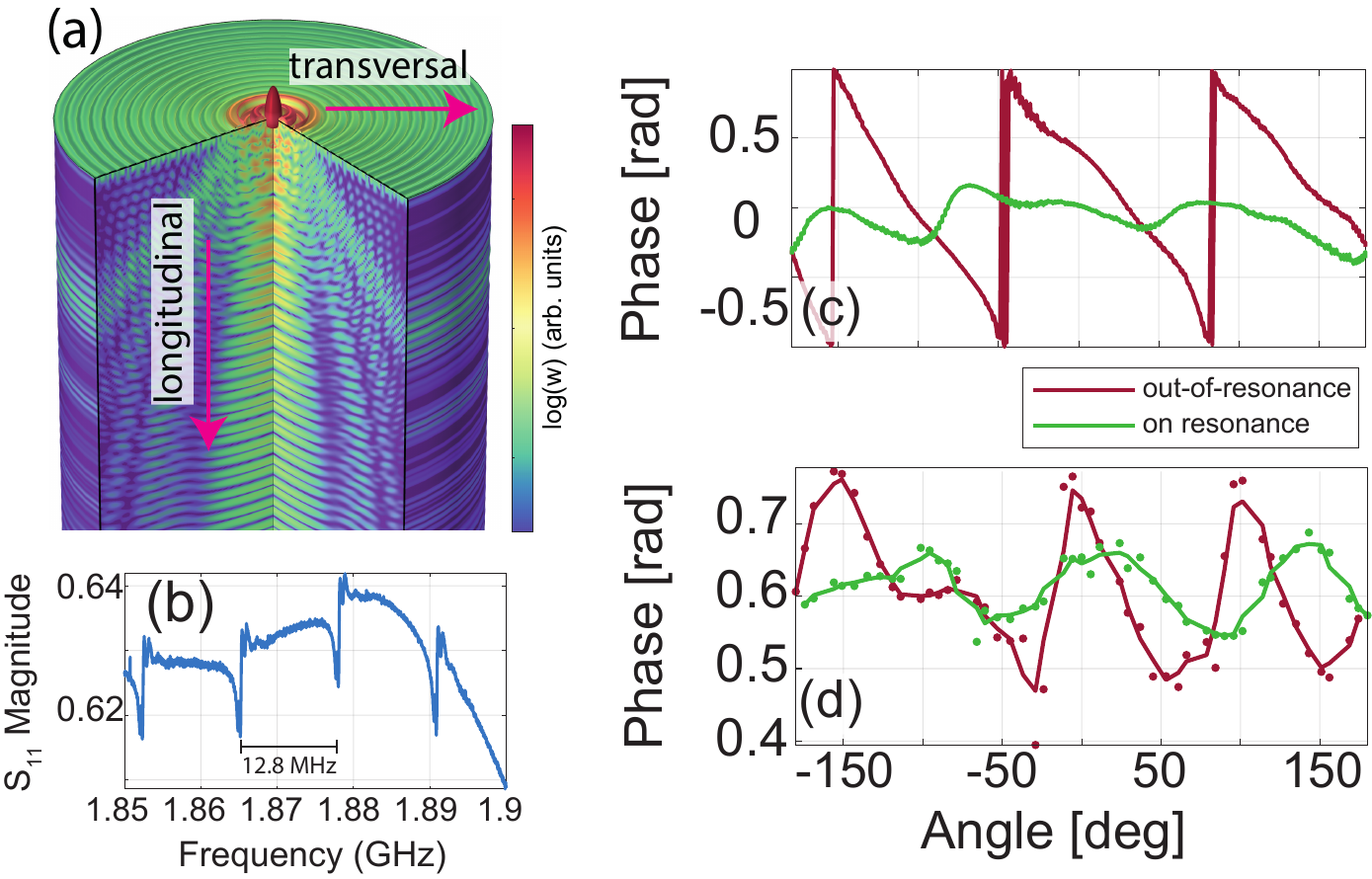}
\caption{\label{fig:2} {\bf Electrical response} 
(a) Simulated mechanical displacement (log-scale) originating from a circular BAWR in a 2D, azimuthally symmetric cell. 
(b) Experimental electrical reflection (corresponding to the  $S_{11}$ rf-scattering parameters) of the BAWR in a limited frequeny range, highlighting the 12.8~MHz separated dips corresponding to substrate resonances. (c) Simulated and  (d) experimental phase profile around the path $\gamma_c$ of Fig. \ref{fig:1}, evaluated precisely at a substrate resonance and far from it. The dots are experimental points, with the solid line representing an average. The phase excursion decreases on the resonance frequency in both simulation and experiment.}
\end{figure}

Interestingly, the observed vortex has a $C_3$ symmetry (i.e., a threefold periodicity along $\theta$, see the dashed lines in Fig.~\ref{fig:1}(e)-(g) as guides for the eyes). This is linked to the presence of a background signal coming from the combination of the transversal and longitudinal components of the waves generated in our device. The waves emitted by a standard BAWRs have, in fact, a strong longitudinal component along with a weaker transversal one. The latter is responsible for the $\theta-\rho$ plane wavefront  engineered for the vortex creation (cf.~simulation of Fig. \ref{fig:2}(a) for a circular BAWR). 
The longitudinal component becomes prominent for frequencies resonating with the substrate thickness: multiple reflections at the substrate boundary usually result in sharp resonances, which can be clearly identified in the electrical signal S-parameter reflection ($S_{11}$). Those have been observed also in our device, as illustrated in  Fig.~\ref{fig:2}(b) and correspond to maxima of the interferometric signal. The peak-to-peak separation, of about 12.8~MHz, corresponds to the frequency spacing between longitudinal acoustic modes with a velocity of $10943\pm552\;m/s$ in the acoustic cavity defined by the boundaries of the c-oriented sapphire substrate with a nominal thickness of 425~$\mu m$. The estimated value is within the value reported for picosecond ultrasonic cryogenic measurements \cite{Hao2001}. 
While the detailed description of the wave components is out of the scope of this manuscript and will be reported somewhere else, here we limit ourselves to mentioning that both longitudinal and transversal components are pervasively present at every excitation frequency, with a dominance of the former one at frequencies corresponding to the substrate resonances.\\

The longitudinal wave component acts as an oscillating background to the engineered planar wavefront which creates the vortex. A detailed model is reported in the Supplementary Note 2, here we just observe that the background modifies both the functional shape of the amplitude and phase signals. The phase excursion evaluated in a path around the vortex reduces with increasing background signal; the phase itself becomes a smoothly oscillating function, whereas it was linearly growing from $0$ to $2\pi\ell$ for a pure vortex of topological charge $\ell$. The amplitude evaluated along the same path undergoes a similar modification, becoming an oscillating function with a number of oscillations equal to $\ell$ in one revolution. This effect was numerically verified employing 3D simulations in a frequency range around the one used to characterize the vortex in Fig.~\ref{fig:1} .
The phase evaluated within the path $\gamma_c$ of Fig \ref{fig:1}  for a frequency with negligible background contribution, i.e. far from the longitudinal wave resonances, shows the expected roughly linear trend (red line in Fig. \ref{fig:2} (c)), with a total $2\pi\times 3$ accumulated phase, thus evidencing the presence of a vortex with $\ell=3$. Conversely, the phase evaluated precisely on resonance shows a decreased excursion range while retaining the presence of the 3 peaks as a signature of the vortex $\ell$ (green line in Fig. \ref{fig:2} (c)).   The same qualitative behaviour has been observed in the phase extracted from the experimental maps: as shown in Fig. \ref{fig:2} (d), the phase excursion is larger when the driving frequency is far from resonance (red curve), while it appears reduced for perfect resonant drive (green curve).  Both dataset show the signature 3 peaks indicating the topological charge; the lack of a quantitative agreement with the simulation has to be searched in the different substrate thickness used in the simulations, which modifies the weights of the background signal.  

\section{Vortex characterization}

The possibility of generating acousto-optical vortices with different topological charges is highly desirable, enabling applications such as AM-multiplexed telecommunication \cite{wang2012,ren2016,xie2018} and topological control of light \cite{shen2019,kim2020} and vibrations \cite{zhang2018,xue2022}.\\

The vortex phase profile can be controlled by engineering the radial pitch of the spiral arm, which defines the top contact shape. This is parameterized by the ratio $\eta=g/\lambda$, where $\lambda$ is the acoustic wavelength and $g$ the spiral spoke length, see Eq. (\ref{eq:spiral}). A vortex with a certain $\ell$ can be then be obtained by employing a  spiral-shaped BAWR with $\eta=\ell$.
This was verified by performing FEM 3D simulations at a fixed frequency and varying $g$, which yield indeed  vortices with $\ell$ ranging 0 to 4 for a fixed driving frequency of 1.134 GHz, in agreement with one would expect by evaluating $\eta$ (cf.  Supplementary Note~3).

The geometrical modification translates in a smooth monotonic variation of the calculated out-of-plane angular momentum, suggesting the presence of very broad and overlapping resonances, each of them centered at $\eta=\ell$. This fact enables the creation of rotating fields in a continuous frequency range and facilitates broad-band modulation in our excitation range from 0.5 to 7 GHz.   
Broadband operation is granted by the specific geometry of our device, with a solid contact whose outermost rim is carved like a spiral; properly shaped BAWR, including multiple metallic fingers arranged in a spiral guise, would lead to sharper resonances and different regime of operations. Furthermore, our simulation excludes the creation of fractionally-charged vortices, which can be observed in more complex configurations, see for example \cite{zhang2022}.\\

\begin{figure}[t!]
\includegraphics[width=0.45 \textwidth]{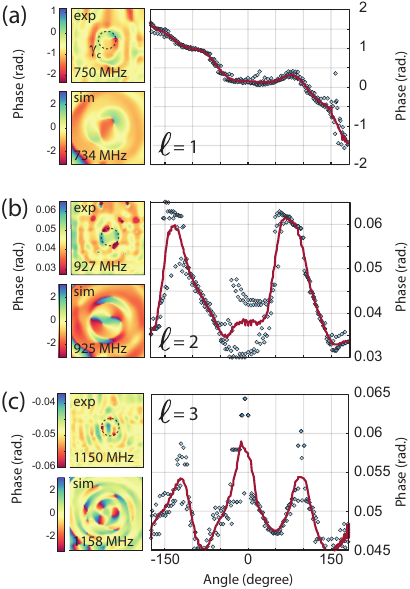}
\caption{\label{fig:3}{\bf Electrically tunable vortex topological charge.} 
(a) Phase profile around a vortex with topological charge $\ell=1$ generated by driving the BAWR at 750~MHz. The upper and lower  left insets show the measured and simulated phase profiles. The dashed circle in the upper left inset shows the circulation path for the determination of the phase profile.
(b)-(c) Corresponding data for vortex with $\ell=2$ and $\ell=3$ excited at 927~MHz and 1150~MHz, respectively. }
\end{figure}
 
In a more practical approach, the vortex topological charge can be tuned by keeping the BAWR geometry (and, thus,  $g$) and simply changing the driving frequency (and thus the acoustic wavelength $\lambda$). We verified this type of frequency-based $\ell$ tuning by experimentally evaluating vortex maps for selected frequencies in the range from 0.5 to 1.5 GHz.
For the sake of measurement speed, all chosen frequencies were tuned to the strongest interferometric signal, which corresponded to a longitudinal BAW resonance, see Fig. \ref{fig:2} (b). Even in presence of strong background signals, we can identify and recognize the vortex topology by looking at the number of phase oscillations in the path $\gamma_c$ around the vortex center, as previously discussed. 
Representative measurements are reported in Fig.~\ref{fig:3}. The experimental phase maps (\textbf{exp}) are compared with the simulated ones (\textbf{sim}), showing a good qualitative agreement between to two for all driving frequencies. Following the previous discussion, we evaluated the experimental phase profile along $\gamma_c$, which has been indicated with a dashed line. As expected, we found a monotonic phase change at 734 MHz, indicating a vortex with $\ell=1$, see Fig. \ref{fig:3} (a). At higher frequencies, we found an increasing number of oscillations, namely two and three at 925 MHz and 1158 MHz, respectively. This indicates the presence of vortices with $\ell=2$ and $\ell=3$, respectively. Note the very different phase range in the experiment, as expected by the increasing strength of the background longitudinal signal with frequency. The latter is compatible with the fact that we are exploring the low-frequency end of the BAWR generation band,  which is centered around 4~GHz and  roughly spans from 0.5 GHz to 7 GH (More details can be found in the Supplementary Note~1).
The spatial resolution of the interferometric setup (of approx. 1~$\mu$m) limits the largest $\ell$ to 4 as measured  at 1325 MHz, see Supplementary Note~4.\\ 
Further insights on the expected maximum topological charge has been gained by running simplified FEM simulations, where we considered a purely mechanical model with a single boundary excitation in the shape of the spiral contact of the BAWR. Despite the rough simplifications, the model well reproduces the main features observed in the experiment, while allowing us to run simulations at an increased spatial resolution. This translated to the observation of a vortex with $\ell=13$ at 5 GHz excitation frequency, which represents a lower bound for the device capabilities.\\
 
Different experimental approaches could be employed for mapping acoustic vortices at very large frequency, such as acoustic atomic force microscopy \cite{Pitanti2023}; alternatively, it is possible to get access to higher $\ell s$ within the same frequency range by realizing more complex spiral contacts. A natural choice would be to consider Archimedean spirals with multiple spokes, which can be introduced by a simple modification of Eq. (\ref{eq:spiral}):
\begin{equation}\label{eq:spiralM}
Sp(\theta) = R_0 + \frac{\mod(M\theta,2\pi)}{2\pi}\cdot g
\end{equation}
where $\mod$ is the modulo function and $M$ is an integer number defining the number of spokes in the spiral. This modification is expected to change the resonant condition for the generation of a vortex with topological charge $\ell$ as:
$\eta' = M\times\eta = \ell$, with $M,\eta\in\mathbb{N}.$

\begin{figure}[b!]
\includegraphics[width=0.4 \textwidth]{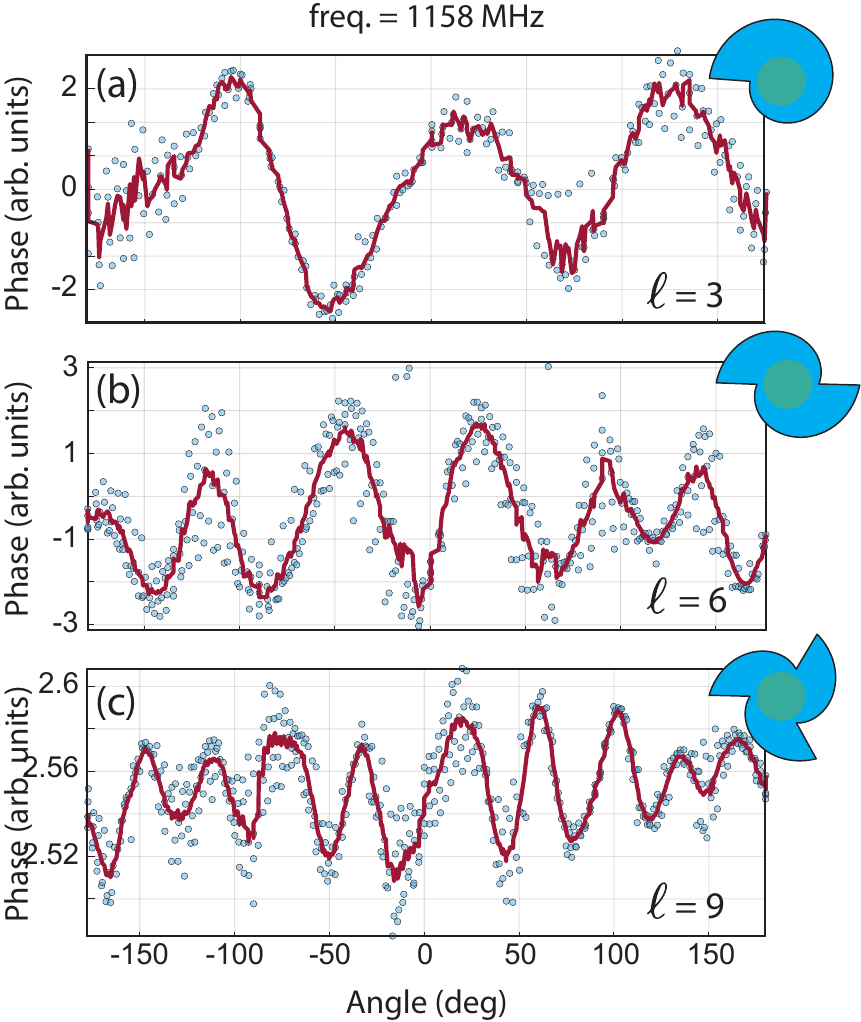}
\caption{\label{fig:4} {\bf High topological charge vortices.} Comparison phase profiles for spiral BAWRs of different $M$s, operated at the same frequency of 1158 MHz. (a): $M=1$ spiral, (b): $M=2$ spiral, (c): $M=3$ spiral. All spiral shapes have been sketched in the upper-right corner of their respective plots. Experimental data points are the blue dots, with the red lines showing their moving mean value.}
\end{figure}

This approach allows a simple scale-up of the topological charge at the same operating frequency by considering $M>1$ in the spiral geometry. Measurements demonstrating this concept are reported in Fig.~\ref{fig:4}. 
Figure~\ref{fig:4}(a)-(c) compare circulation phase profiles recorded by exciting BAWRs with and increasing number of spokes (cf. upper right inset ) at  a fixed  frequency of 1158 MHz. The single-spoke spiral BAWR ($M=1$, cf. Fig.~\ref{fig:4}(a))  generated a vortex with $\ell=3$ for this excitation frequency. For the very same excitation frequency, the double- ($M=2$) and triple-spoke ($M=3$) spiral BAWRs generate, as expected, vortices with $\ell=6$ and $\ell=9$, respectively (cf. Figs.~\ref{fig:4}(b-c)). The scaling-up effect is limited by the device lateral dimension; if $M$ is large enough to reduce the angular spoke-to-spoke distance to less than an acoustic wavelength, it will not allow the formation of the proper planar wavefront which generate the vortex.

\section{Conclusions}

In conclusion, we have experimentally demonstrated the generation of GHz  acousto-optical vortices with tunable topological charge reaching up to 9 using a piezoelectric BAWR. 
The BAWR, which has a compact design based on a double-contacted piezoelectric layer,  generates acoustic vortices propagating into a sapphire substrate. Simultaneously, acousto-optical interaction at the top surface imparts a specific wavefront to light, creating time-varying optical vortex beams  modulated at the mechanical frequency.
Using an interferometric setup we performed coherent characterization of the vortices, detecting both in-phase and quadrature signal and, therefore, access to the phase information. The experimental studies are complemented by modeling and simulation of the acousto-optic vortex beams. 
As a distinguishing feature of our device, we demonstrated the dynamic control of the vortex topological charge by simply tuning the excitation frequency, demonstrating a continuous tuning from $\ell=1$ to $\ell=4$, a maximum experimentally demonstrated $\ell=9$ and estimating a lower bound of $\ell=13$, limited by our simulation platform. 
The operating frequency of acoustic vortex, reaching 7~GHz (with the potential of reaching 20~GHz in similar systems \cite{Kuznetsov2021}), is more than one order of magnitude higher than the largest ones reported in the literature (250 MHz \cite{sahely2022}), while the GHz modulation speed of the time-varying optical vortex well compares with other reports of $\ell-$tunable, solid-state generators (with modulation frequency of $\sim MHz$ \cite{jiang2024}).

The operating range makes it suitable for a new generation of hybrid systems, where acoustic waves are combined with photonic and electronics quantum nanodevices sharing similar, micrometric-sized footprints \cite{Gustafsson2014,Forsch2020,Buhler2022}. On a very general ground, the device characteristics can unveil interesting applications in several fields: from angular-momentum based optical communications \cite{shen2019,Wang2022,he2022}, where $\ell$-tuning and non-resonant light interaction are beneficial to applications, to topological acoustics \cite{Guo2022} and acoustic tweezers \cite{Gong2020,Li2021}, where it is possible to achieve chiral-based 4 degree of freedom particle manipulations \cite{li2024} with the added features of operating at GHz frequencies, where accelerations as high as $\sim10^6 g$ have been predicted for microjets in fluids \cite{daru2024}.

Simple design modifications, e.g., by replacing  the full contact BAWR by spiral-like metallic fingers,  would produce sharp resonances at different $\ell$s, introducing angular momentum degree of freedoms to hybrid quantum systems, such as, for instance, optomechanical excitonic polaritons \cite{Santos2023}, where AM quantization has been recently demonstrated in laser stirring experiments \cite{delValle2023,Gnusov2023} as well as  magnons~\cite{Puebla_APL120_220502_22} and magnetic color centers~\cite{PVS338}.

\appendix
\section{Methods}
\subsection{Device fabrication}
Devices have been fabricated starting with a 425~$\mu$m-thick, double-polished sapphire substrate. The BAWR bottom contact has been defined using optical lithography followed by the sputter deposition of a ZnO piezoelectric layer and the photolithographic  lift-off fabrication of the  10/30/10 nm of Ti/Au/Ti top contact. The bottom contact was shaped as a ring with external radius 25 $\mu$m and internal one 10 $\mu$m. The piezoelectric ZnO layer was sputtered at 150$^\circ$ degree with  a  thickness of 700 nm. This layer has been subsequently patterned using optical lithography and reactive ion etching. 
 The specific shape of the Archimedean spiral top contact was designed using $R_0$ =12.5 $\mu$m, $g$ = 16.67 $\mu$m and different $M$s for the devices with increased number of spokes, see Eq. (\ref{eq:spiralM}). 

\subsection{Numerical simulations}
Numerical simulations were performed using a commercial finite-element method solver (Comsol Multiphysics). Fully coupled electrostatic and mechanical differential equation systems have been solved in order to give a radio-frequency voltage as  input and evaluate the excited mechanical displacement. The two different simulations reported in the manuscript considered a 2D model with full azimuthal symmetry, nominal layer thickness for all the employed materials and a circular BAWR with radius  12.5 $\mu$m (Fig. \ref{fig:2}). Full, 3D simulations considered a $M=1$ spiral with the same nominal dimensions as the fabricated device. Due to memory constrains, while the contacts and ZnO layers have been simulated using their nominal geometric values, the sapphire substrate thickness was 25 $\mu$m (Figs. \ref{fig:1} and \ref{fig:3}). While supporting different resonant conditions for the longitudinal substrate waves, this simplification did not change the qualitative results of the simulations, introducing only a possible mismatch due to the slightly different overlap between the vortex field and the longitudinal wave one.

\subsection{Experimental setup}
The acousto-optical device characterization has been performed using a Michelson-Morley interferometer mounted on a movable microscope head. The device was excited using radio-frequency probes coupled to the port 1 of a Vector Network Analyzer (VNA). A continuous-wave laser light at 532 nm was split into two arms, one ending at an oscillating mirror (modulated at low frequency $f_m$) mounted on a piezoelectric actuator (stabilization arm) and to other leading to a microscope head pointing at the device top surface, which focuses the light beam onto a spot size of around 1 $\mu m$. Once the two beams were recombined upon reflection, the interferometric signal was fed to a fast photodetector. The slow monitor signal at $2f_m$ was used to stabilize the interferometer working point by adjusting the position of the mirror actuator in the stabilization arm; conversely, the fast AC signal recorded  by the photodetector was fed to port 2 of the VNA, leading to a spectrally and phase-resolved resolved detection of the surface displacement profile through the evaluation of the S$_{21}$ scattering parameter. 

\section{Acknowledgments}
A.P. acknowledges funding from the Alexander von Humboldt Stiftung through the ``Experienced Researchers'' fellowship program, PVS acknowledges funding from the German DFG (grant \#426728819).
We also thank A. Riaud and M. Baudoin for discussion in the initial phase of these studies.

\section{Author contributions}
A.P. and P.V.S. conceived and designed the experiment. A.P. and I.d.P.E. characterized the device electrically while A.P., N.A., M.M. and P.V.S. performed the optical interferometric measurements. A.P. and P.V.S. discussed the results, carried out the modelling and performed the numerical simulations. A.P. wrote the manuscript with input and discussion from all the authors.


\bibliography{vortex_paper}

\end{document}